
\input lanlmac
\input tables 

\def\ie{{\it i.e.},$\ $}
\def\l{\left}
\def\r{\right}
\def\eqs#1#2{\eqn #1 {\eqalign{#2}}}

\def\hf{{1 \over 2}}
\def\qt{{1 \over 4}}
\def\Proj{{\cal P}}
\def\d{\partial}

\def\cd{\nabla}
\def\sitarel#1#2{\mathrel{\mathop{\kern0pt #1}\limits_{#2}}}
\def\inbar{\,\vrule height1.5ex width.4pt depth0pt}
\def\IB{\relax{\rm I\kern-.18em B}}
\def\IC{\relax\hbox{$\inbar\kern-.3em{\rm C}$}}
\def\ID{\relax{\rm I\kern-.18em D}}
\def\IE{\relax{\rm I\kern-.18em E}}
\def\IF{\relax{\rm I\kern-.18em F}}
\def\IG{\relax\hbox{$\inbar\kern-.3em{\rm G}$}}
\def\IH{\relax{\rm I\kern-.18em H}}
\def\II{\relax{\rm I\kern-.18em I}}
\def\IK{\relax{\rm I\kern-.18em K}}
\def\IL{\relax{\rm I\kern-.18em L}}
\def\IM{\relax{\rm I\kern-.18em M}}
\def\IN{\relax{\rm I\kern-.18em N}}
\def\IO{\relax\hbox{$\inbar\kern-.3em{\rm O}$}}
\def\IP{\relax{\rm I\kern-.18em P}}
\def\IQ{\relax\hbox{$\inbar\kern-.3em{\rm Q}$}}
\def\IR{\relax{\rm I\kern-.18em R}}
\font\cmss=cmss10 \font\cmsss=cmss10 at 7pt
\def\IZ{\relax\ifmmode\mathchoice
{\hbox{\cmss Z\kern-.4em Z}}{\hbox{\cmss Z\kern-.4em Z}}
{\lower.9pt\hbox{\cmsss Z\kern-.4em Z}}
{\lower1.2pt\hbox{\cmsss Z\kern-.4em Z}}\else{\cmss Z\kern-.4em Z}\fi}
\def\1{{1\hskip -3pt {\rm l}}}



\def\m{\mu}
\def\n{\nu}
\def\p{\rho}
\def\s{\sigma}
\def\dl{\delta}
\def\e{\epsilon}

\def\om{\omega}
\def\G{\Gamma}

\def\CW#1{CW_{#1}(\lambda)}

\def\hpt#1{\href{http://xxx.yukawa.kyoto-u.ac.jp/abs/hep-th/#1}{\tt hep-th/#1}}


\lref\GepWit{
D.~Gepner and E.~Witten,
``String Theory On Group Manifolds,''
Nucl.\ Phys.\ B {\bf 278}, 493 (1986).
}

\lref\Jabbari{
D.~Sadri and M.~M.~Sheikh-Jabbari,
``String theory on parallelizable pp-waves,''
JHEP {\bf 0306}, 005 (2003), 
\hpt{0304169}.
}

\lref\FOF{
J.~Figueroa-O'Farrill, 
``On parallelisable NS-NS backgrounds,'' 
\hpt{0305079}.
}

\lref\AdSS{
J.~M.~Maldacena,
``The large N limit of superconformal field theories and 
supergravity,''
Adv.\ Theor.\ Math.\ Phys.\  {\bf 2}, 231 (1998); 
Int.\ J.\ Theor.\ Phys.\  {\bf 38}, 1113 (1999),
\hpt{9711200}.
}

\lref\AdSSS{
S.~Elitzur, O.~Feinerman, A.~Giveon and D.~Tsabar,
``String theory on $AdS_3\times{S}^3\times{S}^3\times{S}^1$,''
Phys.\ Lett.\ B {\bf 449}, 180 (1999), 
\hpt{9811245}.
}

\lref\PandoZayasBP{
L.~A.~Pando Zayas,
``Ricci-parallelizable spaces in the NS-NS sector,''
\hpt{0007134}.
}

\lref\BCZ{
E.~Braaten, T.~L.~Curtright and C.~K.~Zachos,
``Torsion And Geometrostasis In Nonlinear Sigma Models,''
Nucl.\ Phys.\ B {\bf 260}, 630 (1985).
}

\lref\MetTse{
R.~R.~Metsaev and A.~A.~Tseytlin,
``Order Alpha-Prime (Two Loop) Equivalence Of The String Equations Of Motion 
And The Sigma Model Weyl Invariance Conditions: Dependence On The
Dilaton And The Antisymmetric Tensor,''
Nucl.\ Phys.\ B {\bf 293}, 385 (1987).
}

\lref\BlauJS{
M.~Blau and M.~O'Loughlin,
Nucl.\ Phys.\ B {\bf 654}, 135 (2003),
\hpt{0212135}.
}

\lref\DuffLu{
M.~J.~Duff and J.~X.~Lu,
``Elementary Five-Brane Solutions Of D = 10 Supergravity,''
Nucl.\ Phys.\ B {\bf 354}, 141 (1991).
}

\lref\BoonstraYU{
H.~J.~Boonstra, B.~Peeters and K.~Skenderis,
``Brane intersections, anti-de Sitter spacetimes and dual superconformal 
theories,''
Nucl.\ Phys.\ B {\bf 533}, 127 (1998), 
\hpt{9803231}.
}

\lref\BPS{
H.~J.~Boonstra, B.~Peeters and K.~Skenderis,
``Duality and asymptotic geometries,''
Phys.\ Lett.\ B {\bf 411}, 59 (1997), 
\hpt{9706192}.
}

\lref\CHS{
C.~G.~Callan, J.~A.~Harvey and A.~Strominger,
``World Sheet Approach To Heterotic Instantons And Solitons,''
Nucl.\ Phys.\ B {\bf 359}, 611 (1991). 
}

\lref\Khuri{
R.~R.~Khuri,
``Remarks on String Solitons,'' 
Phys.\ Rev.\ D {\bf 48}, 2947 (1993), 
\hpt{9305143}.
}

\lref\KhuriMG{
R.~R.~Khuri,
``Some Instanton Solutions In String Theory,''
Phys.\ Lett.\ B {\bf 259}, 261 (1991).
}


\Title{                              \vbox{\hbox{UT-03-19}
                                           \hbox{\tt hep-th/0306038}} }
{
\vbox{\centerline{
            Dilatonic Parallelizable NS-NS Backgrounds
}}
}
\vskip .5in

\centerline{
             Teruhiko Kawano  and  Satoshi Yamaguchi
}

\vskip .2in 


\centerline{\sl
               Department of Physics, University of Tokyo
}
\centerline{\sl
                     Hongo, Tokyo 113-0033, Japan
}
\centerline{\tt
                    kawano@hep-th.phys.s.u-tokyo.ac.jp
}
\vskip -0.06in
\centerline{\tt
                    yamaguch@hep-th.phys.s.u-tokyo.ac.jp
}

\vskip 3cm
\noindent

We complete the classification of parallelizable NS-NS backgrounds
in type II supergravity 
by adding the dilatonic case to the result of Figueroa-O'Farrill 
on the non-dilatonic case.
We also study the supersymmetry of these parallelizable backgrounds. 
It is shown that 
all the dilatonic parallelizable backgrounds have sixteen supersymmetries.

\Date{June, 2003}


\newsec{Introduction}
\seclab\intro

The parallelizable pp-wave backgrounds have been recently discussed 
in \Jabbari. 
They are a class of the non-dilatonic NS-NS backgrounds in type II 
supergravity. 
Immediately, Figueroa-O'Farrill has classified all the parallelizable 
NS-NS backgrounds with the dilaton field turned off \FOF. He also 
classified all the ten-dimensional parallelizable NS-NS field configurations 
which do not necessarily satisfy the equations of motion of type II 
supergravity but with a closed torsion in the same paper \FOF. This is 
the classification of the ten-dimensional parallelized Lorentzian 
Lie groups with a bi-invariant metric. 

In this short note, we will turn on the dilaton field to classify all 
the dilatonic parallelizable NS-NS backgrounds in type II supergravity. 
Our strategy is simple: we will find the solution of the dilaton field to 
the equations of motion which is facilitated by the parallelizability 
of the geometry. 
It turns out that all the manifolds in the classification of \FOF\ except for 
$AdS_3\times\IR^7$ and $\CW{4}\times{S}^3\times{S}^3$ can be dilatonic 
parallelizable NS-NS backgrounds and are invariant under half of the 
supersymmetries. The result in this note together with the classification 
in \FOF\ are summarized in Table 1. 
To our knowledge, among the dilatonic backgrounds in Table 1, it seems that 
the case $\IR^{1,1}\times{SU}(3)$ haven't appeared in the literature. 

The parallelizable NS-NS backgrounds are a simple class of curved spacetimes
\refs{\BCZ,\MetTse,\KhuriMG}. 
The propagation of strings on them can be described by the
conformal field theory, in particular by the Wess-Zumino-Novikov-Witten models. 
Some backgrounds are interpreted as the near horizon geometry of
the (intersecting) NS5-branes and/or NS1-branes.
However, not all the backgrounds have been given any interpretation
in terms of the brane configuration. 
Thus, the above $\IR^{1,1}\times{SU}(3)$ might deserve further 
study. Since the NS-NS part of the supergravity action in this note 
is the same as the bosonic action of heterotic supergravity with no Yang-Mills 
background fields, one may be able to generalize the results of this note 
as well as \FOF\ to the heterotic parallelizable backgrounds. 

This note is organized as follows: in section 2, imposing 
the parallelizability on the metric and the NS three-form 
in the equations of motion of type II supergravity, we find a system of 
equations which the dilaton field has to satisfy. 
In section 3, the number of the supersymmetries is 
examined for the dilatonic solutions. The main results of this note are 
summarized in section 4. 

\newsec{Dilatonic Parallelizable NS-NS Backgrounds}
\seclab\second

The NS-NS part of the ten-dimensional type II supergravity action is 
given by 
\eqn\action{
S_{\rm NSNS}={1\over{l}^8_p}\int d^{10}x\sqrt{-g}e^{-2\phi}
\l[R+4\cd_\m\phi\cd^\m\phi-{1\over12}H_{\m\n\p}H^{\m\n\p}\r],
}
yielding the equations of motion 
\eqs{\eom}{
&\cd_\p\l(e^{-2\phi}H^{\p\m\n}\r)=0,
\cr
&\cd_\m\cd_\n e^{-2\phi}=e^{-2\phi}
\l(R_{\m\n}+4\cd_\m\phi\cd_\n\phi-\qt H_{\m\p\s}{H_{\n}}^{\p\s}\r),
\cr
&\cd_\p\cd^\p e^{-2\phi}={1\over6}e^{-2\phi}H_{\m\n\p}H^{\m\n\p},
}
where $\cd_\m$ is the covariant derivative 
with respect to the Levi-Civita connection ${\G^\n}_{\m\p}$. 
The field strength $H_{\m\n\p}$ of the NS-NS two-form potential $B_{\m\n}$ 
is given as usual by $H=dB$. Henceforth, we assume that $H_{\m\n\p}$ is 
closed, \ie $dH=0$. 

Imposing on the metric $g_{\m\n}$ and the field strength $H_{\m\n\p}$ 
the Ricci-parallelizability 
\eqs{\RicciParallel}{
&R_{\m\n}-\qt H_{\m\p\s}{H_{\n}}^{\p\s}=0,
\cr
&\cd_\p H^{\p\m\n}=0,
}
one find that the equations of motion \eom\ are reduced to 
\eqs{\dilaton}{
&H^{\m\n\p}\cd_\p e^{-2\phi}=0,
\cr
&\cd_\m\cd_\n\phi=0,
\cr
&\cd_\p\phi\cd^\p\phi={1\over24}H_{\m\n\p}H^{\m\n\p}.
}
It follows from the above equations \dilaton\ that non-dilatonic background 
solutions are viable, if $H_{\m\n\p}H^{\m\n\p}$ is vanishing, or equivalently 
if the scalar curvature $R$ is zero. This is the case for the parallelizable 
pp-waves \refs{\Jabbari,\FOF}. Instead, if $H_{\m\n\p}H^{\m\n\p}$ is a 
non-zero constant, one can easily solve the equations \dilaton\ with 
the linear dilaton field $\phi$ in the direction transverse to $H_{\m\n\p}$.
The purpose of this note is to consider the latter case in detail. 

In \FOF, Figueroa-O'Farrill have given the classification of not only 
the non-dilatonic parallelizable solutions of the supergravity, but also 
ten-dimensional parallelizable spacetimes, as in Table 1. 
Since the scalar curvature of the latter case is non-zero, 
with the help of the above linear dilaton field, 
all the members of the latter case, except for the two 
$AdS_3\times\IR^7$ and $\CW{4}\times{S}^3\times{S}^3$  
become classical solutions of type II supergravity. 

Let us look at the dilaton gradient more closely.
According to \refs{\FOF}, parallelizable
spacetimes are composed of some of the Minkowski space, a flat space, 
$\CW{2n}\ (n=2,\cdots,5)$, $AdS_3$,
$S^3$, and SU(3) (See table 1). These components are classified into 
the following three classes: 

1. simple groups: $AdS_3$, $S^3$, and SU(3)

2. pp-waves: $\CW{2n}$. 

3. flat spacetimes : $\IR^{1,m},\ \IR^{m}$

First, as for the simple group manifolds, since the torsion of 
the parallelizing connection is proportional to the structure constants 
of the groups, one cannot introduce the dilaton
gradient tangent to the above group manifolds, 
because of the first equation of \dilaton .

Secondly, let us turn to the pp-waves $\CW{2n}$. The metric of this case 
is written \refs{\Jabbari,\FOF} as 
\eqs{\ppwavemetric}{
&ds^2=-2dudv-\sum^{2n-2}_{i,j=1}\mu_{ij}x^{i}x^{j}du^2
+\sum^{2n-2}_{i=1}dx^{i}dx^{i},\quad 
\mu_{ij}=\qt\sum^{2n-2}_{k=1}h_{ik}h_{jk},
\cr
&H=\hf\sum^{2n-2}_{i,j=1}h_{ij}\,dx^{i}\wedge dx^{j}\wedge du. 
}
The matrix $h_{ij}$ is assumed to be invertible. Then, the first equation of
\dilaton\ implies that the dilaton gradient can be introduced
in the $u$ direction. Moreover, the dilaton must be linear in $u$, because of
the second equation of \dilaton. Note that this linear dilaton in the
$u$ direction doesn't contribute to the third equation of \dilaton, since
the $u$ direction is light-like.

Finally, we consider the dilaton field in the flat spacetime. 
The second equation of \dilaton\ requires that the dilaton field be linear 
in the coordinates. If the dilaton gradient is 
time-like, the left-hand side of the third equation of \dilaton\ is 
negative. $H_{\m\n\p}H^{\m\n\p}$ however never becomes negative
in this case since 
the time-like componet of $H_{\m\n\p}$ is zero in the flat spacetime. 
As a result, the third equation of \dilaton\ cannot be satisfied
with the time-like dilaton gradient.
On the other hand, the null dilaton gradient doesn't contribute to 
$\cd_\p\phi\cd^\p\phi$ 
in the third equation of \dilaton\ and can be turned on. 
The most interesting
case is the space-like linear dilaton. In this case, the dilaton field 
can balance the contribution from $H_{\m\n\p}H^{\m\n\p}$ in the third 
equation of \dilaton\ to give the solutions.

In summary, the dilaton can be introduced only linearly in
the $u$ direction of $\CW{2n}$, in the light-like direction of $\IR^{1,n}$, 
and in the space-like of $\IR^n$ as well as $\IR^{1,n}$. 

\vskip 12pt
\vbox{\hbox{
\thicksize=1.5pt
\begintable
\tstrut 
spacetime| non-dilatonic SUSYs | dilatonic SUSYs
\cr
 $AdS_3\times S^3 \times S^3 \times \IR$ | 16 | 16 
\nr
 $AdS_3\times S^3 \times \IR^4$ | 16 | 16 
\nr
 $AdS_3\times \IR^7$ | $\times$ | $\times$
\nr
 $\CW{10}$ | 16, 18(A), 20, 22(A), 24(B), 28(B) | 16
\nr
 $\CW{8} \times \IR^{2}$ | 16, 20 | 16 
\nr
 $\CW{6} \times S^3 \times \IR$ | $\times$ | 16 
\nr
 $\CW{6} \times \IR^{4}$  | 16, 24  | 16
\nr
 $\CW{4} \times S^3 \times S^3$ | $\times$ | $\times$
\nr
 $\CW{4} \times S^3 \times \IR^3$ | $\times$ | 16
\nr
 $\CW{4} \times \IR^6$ | 16 | 16
\nr
 $\IR^{1,1} \times SU(3)$ | $\times$ | 16
\nr
 $\IR^{1,0} \times S^3 \times S^3 \times S^3$ | $\times$ | $\times$
\nr
 $\IR^{1,3} \times S^3 \times S^3$ | $\times$ | 16
\nr
 $\IR^{1,6} \times S^3 $ | $\times$ | 16
\nr
 $\IR^{1,9}$ | 32 | 16
\endtable
}
\vskip 4pt
\hskip 0.8cm {\it \hbox{Table 1: 
the classification of 
ten-dimensional Lorentzian parallelizable}
\vskip -2pt
\hskip 2.1cm\hbox{spaces in \FOF. 
(A) and (B) indicate the cases occurring only in} 
\vskip -2pt
\hskip 2.1cm\hbox{type IIA and IIB supergravity, respectively. 
The cross $\times$ indicates }
\vskip -2pt
\hskip 2.1cm\hbox{that the case in question is not the supergravity background.}
}}

\newsec{Supersymmetry of the Dilatonic Parallelizable Backgrounds}
\seclab\third

The supersymmetry transformation of type II supergravity with 
no R-R background fields is given by 
\eqs{\IIA}{
&\dl\psi_\m
=\l[\d_\m+\qt{\om_\m}^{ab}\G_{ab}+{1\over8}H_{\m ab}\G^{11}\G^{ab}\r]\e,
\cr
&\dl\lambda=\l[\G^\m\cd_\m\phi+{1\over12}\G^{abc}H_{abc}\G^{11}\r]\e,
}
for type IIA supergravity, and 
\eqs{\IIB}{
&\dl\psi_\m
=\l[\d_\m+\qt{\om_\m}^{ab}\G_{ab}+{1\over8}H_{\m ab}\G^{ab}\s_3\r]\e,
\cr
&\dl\lambda=\l[\G^\m\cd_\m\phi+{1\over12}\G^{abc}H_{abc}\s_3\r]\e,
}
for type IIB supergravity, where $\s_3$ is the generator of 
the $SL(2,\IR)$ symmetry of type IIB supergravity.


The supersymmetry transformation for the gravitinos $\psi$ imposes on 
the Killing spinors either of the conditions 
\eqn\gravitino{
\l[\d_\m+\qt{\om_\m}^{ab}\G_{ab}\pm{1\over8}H_{\m ab}\G^{ab}\r]\e=0.
}
Since the parallelizability implies that the curvature tensor including 
the torsion vanishes, it means that the holonomy is locally trivial 
\refs{\Jabbari,\FOF}. In this note, since our attention will be restricted to 
simply connected spaces, the locally trivial holonomy guarantees the existence 
of the solutions to \gravitino. Thus, one concludes that the condition 
\gravitino\ impose no conditions on the Killing spinors, 
as in the non-dilatonic case 
\foot{More careful examination is needed to find the solutions to \gravitino\ 
for the discrete quotients of the spaces under consideration. A simple example 
is provided by the $S^1$ compactified parallelizable pp-waves \Jabbari, where 
the condition \gravitino\ actually reduces the number of the Killing spinors.}\refs{\Jabbari,\FOF}.


We proceed to the supersymmetry transformation for the dilatinos $\lambda$, 
which imposes on the Killing spinors either of the conditions
\eqn\dilatino{
\l(\G^\m\cd_\m\phi\pm{1\over12}\G^{abc}H_{abc}\r)\e=0. 
}
Since the parallelizability means that all the members listed in Table 1 
are group manifolds, 
the field strength $H_{\m\n\p}$ of the parallelizing connections 
is proportional to the structure constants of the corresponding groups. 
Therefore, by making use of the Jacobi identity of the structure constants, 
one find that 
\eqn\HH{
\l(\G^{abc}H_{abc}\r)^2=-6\, H_{\m\n\p}H^{\m\n\p}. 
}
First, we consider the case with non-zero $\cd_\m\phi\cd^\m\phi$. 
Since $\cd_\m\phi$ is transverse to $H_{\m\n\p}$, by multiplying 
$\G^\m\cd_\m\phi$ with both sides of \dilatino, 
one obtain 
\eqn\halfspinor{
0=\l({\cd_\p\phi\cd^\p\phi}\pm{1\over12}
\G^{\s\m\n\lambda}\cd_\s\phi\cdot{H}_{\m\n\lambda}\r)\e
\equiv 2\l({\cd_\p\phi\cd^\p\phi}\r)\,\Proj_{\pm}\e,
}
where the operators $\Proj_\pm$ 
are the projectors satisfying that 
$(\Proj_{\pm})^2=\Proj_{\pm}$ and that $\Proj_++\Proj_-=1$. 
These projectors also satisfy ${\rm Tr} \left[{1 \over 2}(1\pm 
\Gamma^{11})\Proj_{\pm}\right]=8$, where the first signs $\pm$ are not related 
to the second ones. 
The projectors $\Proj_{\pm}$ guarantee that half of the supersymmetries 
survive the condition \halfspinor.

Next, we turn to the case of the null linear dilatonic background, \ie 
$\cd_{\m}\phi \cd^{\m}\phi=0$. This type of the backgrounds
is $\CW{2n}\times \IR^{10-2n}$ or $\IR^{1,9}$, where 
only the non-zero component $\cd_{u}\phi$ is constant.
The dilatino condition \dilatino\ reads
\eqn\nulldilaton{
\l(\cd_{u}\phi \pm {1\over 4}\G^{ij}h_{ij}\r)\G^{+}\e=0.
}
Here, the matrix $\G^{ij}h_{ij}$ is an anti-hermitian matrix
and its eigenvalues are pure imaginary or zero. This means that the matrix
$\l(\cd_{u}\phi \pm (1/4)\G^{ij}h_{ij}\r)$ 
is invertible. It follows from this fact that the ones satisfying that 
$\G^{+}\e=0$ are the solutions to \nulldilaton: 
the backgrounds in this case preserve sixteen supersymmetries.

Thus, all the dilatonic solutions are invariant under sixteen 
supersymmetries. 

\newsec{Summary}
\seclab\summary

The main result of this paper is summarized in Table 1. 
In this section, we will try to elaborate a little more on it. 

\subsec{$AdS_3\times{M}_7$}

The anti-de Sitter space $AdS_3$ can be parallelizable. 
Since the scalar curvature of the anti-de Sitter space is negative, 
the case $M_7=\IR^7$ cannot be the background of the supergravity. 
In this case, $H_{\m\n\p}H^{\m\n\p}$ is also negative. As is seen from 
the last equation of \dilaton, the dilaton field $\phi$ would be imaginary. 
In the two cases of $M_7=S^3\times S^3\times\IR$ \refs{\BoonstraYU,\AdSSS}  
and $M_7=S^3\times\IR^4$ \refs{\BPS,\AdSS},
since $S^3\simeq{SU}(2)$ is parallelizable and the curvature of $S^3$ is 
positive, one can adjust the radii of the three-spheres to set the total 
scalar curvature to zero. Then, they become the non-dilatonic backgrounds of 
the supergravity. Even if the contribution of the three-spheres to 
the total scalar curvature overcomes the one of the $AdS_3$ space, 
the dilatonic solutions are viable with the space-like linear dilaton field 
$\phi$. Actually, in the case of $M_7=S^3\times S^3\times\IR$,
the relation of the radii and the dilaton becomes
\eqn\radii{
\cd_{\m}\phi\cd^{\m}\phi+{1 \over L_{0}^2}-{1 \over L_{1}^2}-{1 \over L_{2}^2}=0,
}
where $L_0$ is the radius of $AdS_3$, and $L_1,L_2$ are the radii of
two $S^3$. The relation for the case with $M_7=S^3\times \IR^4$
is obtained by the limit $L_2\to \infty$ in \radii.
The parallelizing connections of $AdS_3\simeq{SL}(2,\IR)$ and 
$S^3\simeq{SU}(2)$ include the torsion proportional to the structure 
constant of the corresponding groups. Therefore, 
as is seen from the equations in \dilaton, the dilaton field $\phi$ 
can linearly depend only on the coordinate in the direction normal to 
$AdS_3$ and $S^3$. 
Whether they are dilatonic or not, they are invariant 
under half of the supersymmetries, as explained above. 

\subsec{$\CW{2n}\times{M}_{10-2n}$}

The field strength $H_{\m\n\p}$ of 
the parallelizable pp-waves $\CW{2n}$ doesn't contribute to 
$H_{\m\n\p}H^{\m\n\p}$. Indeed, 
the parallelizable pp-wave backgrounds $\CW{2n}$
are given \refs{\Jabbari,\FOF} by \ppwavemetric. 
In the case of $\CW{4}\times{S}^3\times\IR^{3}$ and 
$\CW{6}\times{S}^3\times\IR$, 
thus, the only contribution to $H_{\m\n\p}H^{\m\n\p}$ comes from the 
three-sphere $S^3$. The first equation of \dilaton\ allows 
the dilaton gradient $\cd_\m\phi$ to have a non-zero $u$-component and 
non-zero components in the spatial direction normal to $\CW{2n}$ and $S^3$.
However, $\cd_u\phi$ doesn't contribute to $\cd_\p\phi\cd^\p\phi$ 
while $\cd_v\phi$ is kept zero. This is the reason why 
$\CW{4}\times{S}^3\times{S}^3$ isn't the NS-NS background of 
the supergravity. In the case in question, there is the factor $\IR$, 
in the direction of which the dilaton field can be turned on to yield 
the solution to \dilaton. 

For the case $M_{10-2n}=\IR^{10-2n}$, one can find the non-dilatonic 
backgrounds \refs{\Jabbari,\FOF}, where the diverse numbers of supersymmetries 
are available. Furthermore, by turning on the null linear dilaton field, 
one obtain the dilatonic solutions, which are invariant under half of the 
supersymmetries. 

\subsec{$\IR^{1,n}\times{M}_{9-n}$}

As is seen from Table 1, $M_{9-n}$ can be $SU(3)$, $S^3\times{S}^3$ \Khuri, 
and $S^3$ \refs{\DuffLu,\CHS} to yield the supergravity backgrounds. Obviously, 
the ten-dimensional Minkowski spacetime $\IR^{1,9}$ is also the solution. 
All in the former case have the positive scalar curvature. 
The field strength $H_{\m\n\p}$ given by the parallelizing connection 
is proportional to the structure constant of the respective groups. 
Therefore, the space-like linear dilaton is needed to be turned on 
in the direction tangent to $\IR^{n}$. Although the ordinary Minkowski 
spacetime $\IR^{1,9}$ is the supergravity solution with all the 
thirty-two supersymmetries, 
one can see that the Minkowski spacetime $\IR^{1,9}$ with the null 
linear dilaton is also the solution with half of the supersymmetries.


\vskip 0.5in
\centerline{{\bf Acknowledgements}}
We would like to thank Yosuke Imamura, Yu Nakayama, Yuji Tachikawa, and 
Hiromitsu Takayanagi for discussions. We are also grateful to Jos\'e 
Figueroa-O'Farrill for pointing out the missing manifold in the original 
Table 1. 
The work of S.Y. was supported in part by JSPS Research Fellowships
for Young Scientists.

\bigskip\bigskip
\noindent
{\bf Note added:} after publishing this paper in Phys.~Lett.~B568 (2003) 
78-82, we were informed from Jos\'e Figueroa-O'Farrill 
that the parallelizable manifold $\IR^{1,0}\times{S}^3\times{S}^3\times{S}^3$ 
should be included in the original Table 1. 

On this manifold,  it follows from the first equation in \dilaton\ that only 
the time-like dilaton gradient can be turned on, which can only give the 
negative contribution to the left-hand side of the third equation in \dilaton.  
Therefore, since the right-hand side of the same equation is positive, 
the manifold is not a solution to the equations of motion. 
Thus, our result on the classification of the dilatonic parallelizable 
backgrounds remains intact. 


\listrefs

\end